\documentclass[english,aps,twocolumn,superscriptaddress]{revtex4-1}
\usepackage[T1]{fontenc}
\usepackage[latin9]{inputenc}
\usepackage{amstext}
\usepackage{wasysym}
\usepackage{graphicx}
\usepackage{esint}

\makeatletter
\@ifundefined{textcolor}{}
{%
 \definecolor{BLACK}{gray}{0}
 \definecolor{WHITE}{gray}{1}
 \definecolor{RED}{rgb}{1,0,0}
 \definecolor{GREEN}{rgb}{0,1,0}
 \definecolor{BLUE}{rgb}{0,0,1}
 \definecolor{CYAN}{cmyk}{1,0,0,0}
 \definecolor{MAGENTA}{cmyk}{0,1,0,0}
 \definecolor{YELLOW}{cmyk}{0,0,1,0}
}

\makeatother

\usepackage{babel}
\begin{document}

\title{Stiffening of Red Blood Cells Induced by Disordered Cytoskeleton
Structures: A Joint Theory-experiment Study}

\author{Lipeng Lai}

\affiliation{Singapore-MIT Alliance for Research and Technology (SMART) Centre,
Singapore 138602}

\affiliation{Department of Chemistry, Massachusetts Institute of Technology, Cambridge,
USA 02139}

\author{Xiaofeng Xu}

\affiliation{Singapore-MIT Alliance for Research and Technology (SMART) Centre,
Singapore 138602}

\affiliation{NUS Graduate School for Integrative Sciences and Engineering, National
University of Singapore, Singapore 119077 }

\author{Chwee Teck Lim}

\affiliation{Singapore-MIT Alliance for Research and Technology (SMART) Centre,
Singapore 138602}

\affiliation{NUS Graduate School for Integrative Sciences and Engineering, National
University of Singapore, Singapore 119077 }

\affiliation{Nano Biomechanics Laboratory, Department of Biomedical Engineering
and Department of Mechanical Engineering, National University of Singapore,
Singapore 119077}

\affiliation{Mechanobiology Institute, National University of Singapore, Singapore
119077}

\author{Jianshu Cao}
\email{jianshu@mit.edu}
\thanks{Corresponding author}

\affiliation{Singapore-MIT Alliance for Research and Technology (SMART) Centre,
Singapore 138602}
\affiliation{Department of Chemistry, Massachusetts Institute of Technology, Cambridge,
USA 02139}

\begin{abstract}
The functions and elasticities of the cell are largely related to
the structures of the cytoskeletons underlying the lipid bi-layer.
Among various cell types, the Red Blood Cell (RBC) possesses a relatively
simple cytoskeletal structure. Underneath the membrane, the RBC cytoskeleton
takes the form of a two dimensional triangular network, consisting
of nodes of actins (and other proteins) and edges of spectrins. Recent
experiments focusing on the malaria infected RBCs (iRBCs) showed that
there is a correlation between the elongation of spectrins in the
cytoskeletal network and the stiffening of the iRBCs. Here we rationalize
the correlation between these two observations by combining the worm-like
chain (WLC) model for single spectrins and the Effective Medium Theory
(EMT) for the network elasticity. We specifically focus on how the
disorders in the cytoskeletal network affect its macroscopic elasticity.
Analytical and numerical solutions from our model reveal that the
stiffness of the membrane increases with increasing end-to-end distances
of spectrins, but has a non-monotonic dependence on the variance
of the end-to-end distance distributions. These predictions are verified
quantitively by our AFM and micropipette aspiration measurements of
iRBCs. The model may, from a molecular level, provide guidelines for
future identification of new treatment methods for RBC related diseases,
such as malaria infection. 
\end{abstract}
\maketitle

\section{INTRODUCTION}

The mechanical properties of a system are largely dictated by its
structure. The property-structure relationship has been studied extensively
in different fields in physics and engineering(e.g., \cite{discher94, suresh06,broedersz14}, etc.). In recent years, networks
in biological systems have drawn much attention due to their close relationships
to the functions of organisms. Example systems of biopolymer networks
are cytoskeletons in various cells, which can be quasi-1-dimensional
(e.g., axons of neuron cells\cite{xu12,lai14}), 2-dimensional (e.g.,
red blood cells (RBCs)), or 3-dimensional. A biopolymer network can
behave very differently from a network made of synthesized polymers.
Firstly, biopolymer networks can be active with the participation
of ATPs\cite{kim13}. Secondly, the components of a biopolymer network
usually follow the worm-like chain (WLC) behavior, whose elasticity
has an entropic origin and thus a nonlinear dependence on the end-to-end
distance of corresponding bio-filaments.

On the cellular level, the biological functions and behaviors of cells
are related to their mechanical properties\cite{suresh06,huang2014}.
The mechanical properties of cytoskeletons and membranes have been
studied intensively in different scenarios, via experiments (e.g.,
\cite{discher94,lenormand01,hale11}), simulations (e.g., \cite{saxton90,boal94,hansen97,boey98}),
and theoretical modelings (e.g., \cite{nir2007,broedersz14}). In addition,
the membrane of RBCs is also investigated carefully from both biological
and physical perspectives, including the functions of  transmembrane
proteins, and the interactions between protein complexes and the spectrins
in the cytoskeletal network (e.g., \cite{byers85,nir2007}), etc. As an example
to illustrate the relationship between the mechanical properties of
the membranes and the behavior of the cells, earlier studies showed
that the adhesiveness and hence the mobility of RBCs
are strongly affected by the stiffness of the membrane (including
the lipid bilayer and the cytoskeleton)\cite{efremov11,xu13}. Relating
to this paper, recent experiments reveal that the stiffening of RBC
after being infected by malaria parasites \cite{chien1978,nash89,paulitschke1993,glenister2002} correlates
with the structural transformation in the cytoskeleton of the infected RBCs (iRBCs)\cite{shi13}.
A similar correlation was also observed in iRBCs after chloroquine
treatment. In both cases, it is found that, when RBCs were infected
by malaria or iRBCs were treated with chloroquine, the shear modulus
of the membrane increased with time. Meanwhile, the average length
of the spectrins that formed the cytoskeletal network was increased.
It is also noticed that the cytoskeleton mesh became more irregular
with large holes that were absent in the cytoskeleton of normal RBCs,
creating a broad distribution of hole sizes and spectrin lengths.
It is known that the adhesiveness of RBCs is one of the reasons that
leads to the fatality of malaria infection\cite{hughes2010}. Since the stiffness of
RBCs largely affects the adhesiveness of the cells, having a fundamental
understanding of the relationship between the network structure and
the macroscopic elastic properties will provide us with guidances
in potentially discovering new drug targets or treatment methods.

In this study, we focus on rationalizing the correlation between the
two experimental observations mentioned above and investigate how
the structural changes at the molecular level affect the mechanical
properties at the cellular level. We present both numerical solutions
and analytical approximations based on a model combining the worm-like
chain (WLC) description for single filaments and the Effective Medium
Theory (EMT). We use the cytoskeletal network of the RBC, approximated
by a 2-dimensional triangular network, to demonstrate any agreement
between the model and the experiments. This study not only provides
a further understanding of the relationship between structures and
functions, but also provides a critical experimental test of the theoretical
predictions. In addition, because our model is constructed with a
general framework, it is expected that it can also be applied to other
scenarios, such as 3-dimensional networks of the cytoskeletons in
other types of cells, designing of new bio-materials, etc.

\section{METHODS}

\subsection{Experiments}
\subsubsection{Cell culture and enrichment}
The common laboratory P. falciparum 3D7 parasites were used for the study. Parasite iRBCs were cultured {\it{in vitro}} following a conventional protocol\cite{trager1976}. Culture stage synchrony was maintained using sorbitol treatment \cite{ribaut2008} and the early trophozoite stage iRBCs were used for the following experiments.
\subsubsection{AFM experiment}
\paragraph{Sample preparation:} The AFM imaging samples were prepared following similar protocol used previously\cite{shi13}. Briefly, synchronized and MACs enriched \cite{ribaut2008} early trophozoite stage malaria iRBCs after different hours of 1 $\mu M$ chloroquine treatment (0hr, 4hr, 8hr, 16hr and 24hr) were loaded on PHA-E coated cover slips and then incubated for 4 hours to allow sufficient contact time for iRBCs to adhere to the substrate and avoid whole cell detachment in the shear-wash step. 80 ml of 5P8-10 buffer was used to shear-wash the iRBCs adhered to PHA-E coated cover slips at an angle of around 20$^\circ$ by syringe, leaving only the tightly bounded membranes. The tight binding between membrane and substrate also maximally ensured the structure of the cytoskeleton remained {\it in situ}. The cytoplasmic-surface-exposed samples were checked under phase contrast microscope (Olympus X71) and then vacuum dried before imaging using AFM (Fig. S1).

\paragraph{AFM imaging:} AFM imaging was carried on JPK Nanowizard$^{\textregistered}$ AFM with tapping mode (air) using SSS-NCHR AFM tip (NANOSENSORSTM) with tip radius of 2 nm, and Dimension FastScanTM Atomic Force Microscope (Bruker) using Fast Scan AFM tip with tip radius of 5 nm. Images were captured at the resolution of 512 x 512 pixels for 10 $\mu m$ x 10 $\mu m$ or 1 $\mu m$ x 1 $\mu m$ at scan rate of 0.5 to 1 $Hz$ depending on the scan scale and image quality. Around 10 iRBCs were scanned for each chloroquine treatment condition and a well connected spectrin area was selected to represent the cell for spectrin lengths data collection. 10 to 30 spectrin lengths were collected from each cell and averaged to represent the spectrin length of that cell.

\paragraph{Measurement of the spectrin length:}
The spectrin length was measured as described previously\cite{shi13}. As shown in Fig. S2(a) and (b), white lines that were traced along the spectrins between two junctions were measured as the lengths of the spectrins. 

\subsubsection{Micropipette experiment}
Different durations of chloroquine treated malaria iRBCs were prepared at a ratio of 1: 2000 (cell pellet: 1$\%$ BSA solution in PBS). 400-600 $\mu l$ of the diluted sample was pipetted into a home-made micropipette aspiration microscope cell holder which was mounted on the Olympus X71 microscope stage using masking tape. Borosilicate glass tubings (Sutter) were pulled using a Sutter 2000 micropipette puller, then forged and cut into micropipette tip with an inner diameter of 1 -- 2 $\mu m$ using Narishige Microforge. A syringe pump was used to apply a negative suction pressure to aspirate the iRBC into the micropipette. A simple relationship between elongation of cell membrane aspirated into the micropipette and the applied negative pressure is given by \cite{chien1978}:
\begin{equation}\label{eq:micro-pipette}
(\Delta P\times D_p)/\mu = 2.45\times \Delta L_p/D_p
\end{equation}
with $\Delta L_p/D_p > 1$, and $\Delta P$ is the applied negative pressure, $D_p$ is the micropipette inner diameter, $\mu$ is the membrane shear elastic modulus and $\Delta L_p$ is the aspirated length of the cell membrane into the micropipette. Series of images were captured using a CCD camera every second. By measuring the aspirated length and the applied negative pressure, the membrane shear modulus was obtained.

\subsection{Irregular network of worm-like chains (WLC)}

Normal RBC possesses a triangular cytoskeletal network consisting
of nodes of protein complexes and edges of spectrins. In our theoretical study of the RBC cytoskeleton, we regard the network
formed by equilateral triangles as the regular network. We treat this
regular network as our reference state and study how disorders added
to the network affect its elasticity. 

\begin{figure}

\centerline{\includegraphics[scale=0.3]{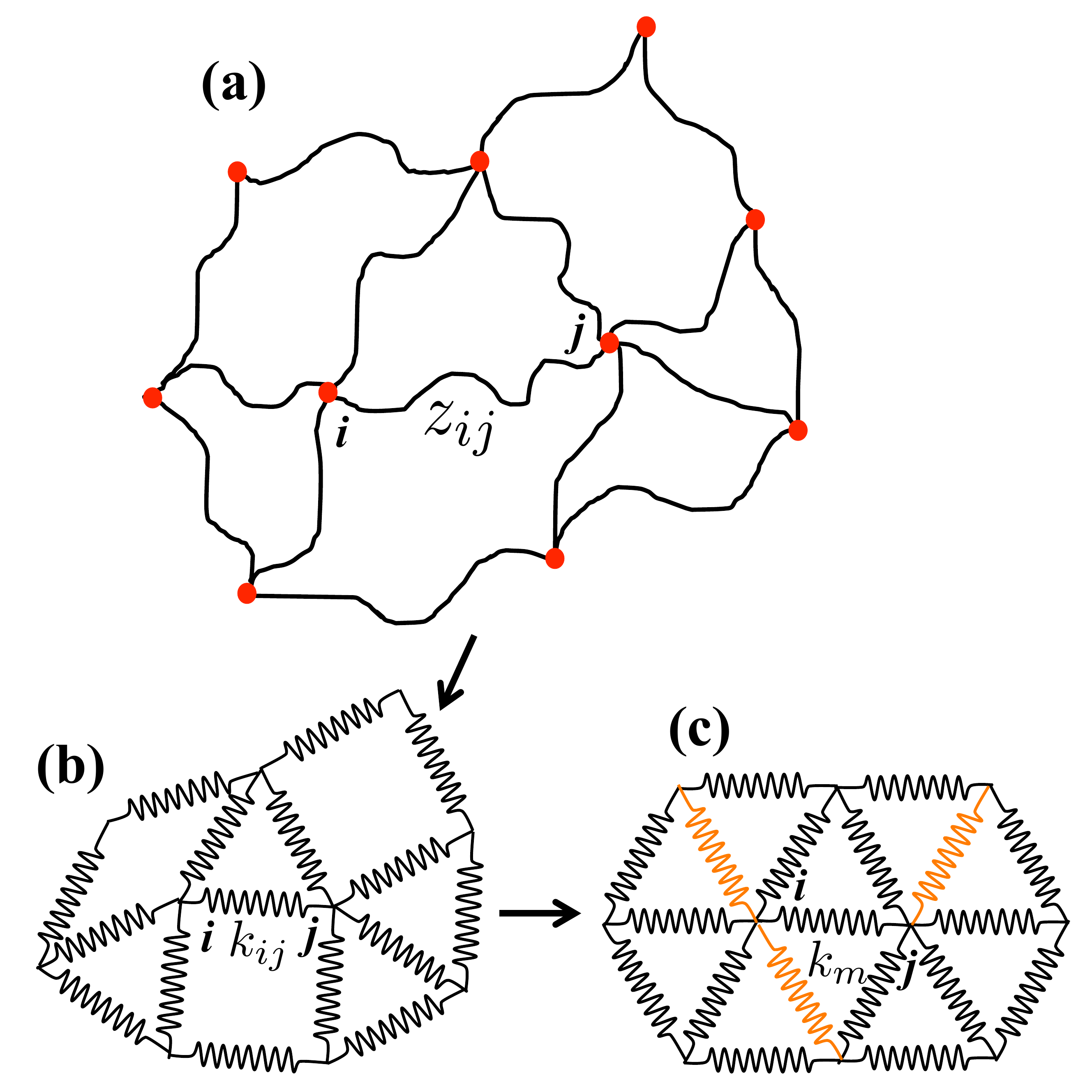}}\protect\caption{\label{fig:Sketch-of-model}Sketches of the model. (a) A sketch of
the 2D triangular network with disorders, resembling the cytoskeleton
structure of the RBC (or iRBC). The lengths $z_{ij}$ between two
nodes are non-uniform and some connections are missing comparing to
a regular network. (b) Based on the WLC model (Eq. \ref{eq:wlc-force-extension},
and Eq. \ref{eq:wlc-effective-spring-constant}), the length variation in
$z_{ij}$ is mapped to the variation in the equivalent spring constant
$k_{ij}$ for small perturbations. (c) An EMT is applied to reduce
the disordered network to a regular network with uniform spring constant
$k_{m}$, from which the macroscopic in-plane shear modulus can be
calculated.}

\end{figure}

Recent experiments \cite{shi13} found that in malaria infected
RBCs (iRBCs) or iRBCs after drug (chloroquine) treatment, two types
of disorders are introduced to the RBC cytoskeletal network (Fig. \ref{fig:Sketch-of-model}). Firstly,
the average end-to-end distance of the spectrins increases, which
is directly measured in our experiment (Fig. \ref{fig:Experimental-results}
(b)). Secondly, the variance of the end-to-end distance distribution
is also much larger than that in the regular triangular network (Fig.
\ref{fig:Experimental-results} (b) and \cite{shi13}), probably due
to the absence of some spectrin links after infection or drug-treatment.
To perform a quantitative investigation of how these two changes affect
the stiffness of the RBC network, we introduce the disorders in terms
of a Gaussian distribution of the spectrins' end-to-end lengths $z$ in the network,
for which the probability density function can be written as:
\begin{equation}
\tilde{{\cal P}}(z)=\frac{C}{\sqrt{2\pi}\sigma}e^{-\frac{(z-z_{0})^{2}}{2\sigma^{2}}}~~0<z<L\label{eq:length-distribution}
\end{equation}
Here, $L$ is the natural length of the spectrins following the WLC
description elaborated later, $z_0$ is the average end-to-end length of the spectrins, $\sigma^2$ is the variance of the end-to-end length distribution, and $C$ is a normalization constant.

Each node in the network formed by protein complexes defines a point
connecting the cytoskeleton and the membrane. In the regular network,
each node is linked by $6$ spectrins, i.e., an out-degree of $6$.
But in reality, some spectrins may lose the connections with the nodes,
which introduces a topological disorder to the network. Here we use
a parameter $p$ to represent the probability with which an edge of
spectrin is present in the regular network (then $1-p$ represents
the probability with which an edge of spectrin is missing from the
regular network). For example, $p=1$ corresponds to the complete
network, and $p=2/3$ corresponds to an average out-degree of $4$.
In our model, we generally treat $p$ as an independent variable.
But it is possible that $p$ can be a function of stresses in the
network as suggested by previous studies focusing on the binding/unbinding
kinetics between ligands and receptors (e.g., \cite{bell78}). This
case is further discussed in the Discussion section.

For individual spectrins, it is well-known that the elasticity has
an entropic origin. Here, we apply the worm-like chain (WLC) model
that is used to study semi-flexible polymers (e.g., \cite{marko95,yang03,xu2014})
to single spectrins, which basically gives us a nonlinear force-extension
curve\cite{marko95}. Comparing to the end-to-end distance of the
spectrins in the network ($\sim50~nm$), the persistence length ($l_{p}\sim10$$nm$)
of the spectrin is very short. Hence we need not consider the effect
of the spectrins' bending rigidity on the network. For a brief review,
in the WLC model, the persistence length $l_{p}$ describes the length
scale, below which the bending energy dominates the thermal excitations.
From the WLC model, the single spectrin follows the force-extension
relation as shown below\cite{marko95}:

\begin{equation}
\beta l_{p}f=\frac{z}{L}+\frac{1}{4(1-z/L)^{2}}-\frac{1}{4}\label{eq:wlc-force-extension}
\end{equation}
where $\beta=1/k_{B}T$ is the Boltzmann factor with $k_{B}$ the
Boltzmann constant and $T$ is the temperature. $z$ is the
end-to-end distance of the spectrin and $L\sim200~nm$ is the natural
length of the spectrin that is treated as in-extensible. With $L\sim 200~nm$ and $l_p\sim 10~nm$, the root mean square end-to-end distance of a spectrin free of stress is about $60 ~nm$, comparable to observed values. In general,
the spectrins in the RBC cytoskeleton has an average
end-to-end distance ranging from about $50~nm$ to $100~nm$. With
small perturbations to the spectrin from its equilibrium position,
we can expand the force-extension relation (eq. (\ref{eq:wlc-force-extension}))
to first order and obtain an effective spring constant $k$ for the
spectrin, which is a function of the end-to-end distance $z$ of the
spectrin at equilibrium. Simply, this extension-related spring constant
$k$ is obtained by taking the derivative of the force with respect
to the extension, which gives us
\begin{equation}
k(z)=\frac{\partial f}{\partial z}=(1+\frac{1}{2}(1-\frac{z}{L})^{-3})/\beta l_{p}L\label{eq:wlc-effective-spring-constant}
\end{equation}

Importantly, we should note here that, even the effective spring constant
$k$ depends on the end-to-end distance $z$ nonlinearly, its effect
in our model is simply to setup an initial equilibrium distribution
of the spring constants across the entire network according to the
distribution of the end-to-end distance of the spectrins. Under this
assumption, we focus on the linear response of the disordered network
(small perturbations), in which case each spectrin is treated as a
Hookean spring with a distinct spring constant and then we apply an
Effective Medium Theory (EMT) to evaluate the elasticity of the disordered
network.

\subsection{Effective medium theory}

To understand how the disorders in the RBC cytoskeleton affect its
stiffness, we adopt an Effective Medium Theory (EMT) approach here.
In previous studies, EMT is used to understand the elastic properties
of different types of disordered networks. Examples include but are
not limited to central-force spring networks \cite{feng85,thorpe90},
2D networks under tension \cite{tang88,boal93,boal98}, networks at
large deformations \cite{sheinman12}, networks formed by filaments
with finite bending rigidities \cite{das07,mao13}, networks with
nonlinear cross-linkers \cite{broedersz09}, etc. Basically, the EMT
maps the disordered network to an equivalent regular network and extract
elastic constants from the regular network. To construct the equivalent
regular network, the springs in the network with distinct spring
constants are replaced by springs with the same spring constant $k_{m}$
self-consistently. The self-consistency requires that the average extra displacement
$\delta u$ caused by this replacement procedure must be $0$
over the entire network ($<\delta u>=0$) \cite{feng85}. It leads
to the following equation for the effective spring constant in the
regular network\cite{feng85}:

\begin{equation}
\int\frac{{\cal P}(k)}{1-\alpha(1-k/k_{m})}dk=1\label{eq:EMT}
\end{equation}
where ${\cal P}(k)$ describes the distribution of the spring constants
in the network. $\alpha$ is related to the topology of and pre-stresses
in the network \cite{tang88} and can be understood as follows: When
two adjacent nodes in the regular network is displaced with respect
to each other, the response of the network can be described by an
effective spring constant $K'=k_{m}/\alpha$, where $0<\alpha<1$
takes into account the contribution of the whole network.
Eq (\ref{eq:EMT}) can be solved analytically for specific probability
density functions ${\cal P}(k)$ and numerically for other forms of
${\cal P}(k)$.

\section{RESULTS}
\subsection{Experimental measurements}

In the cytoskeleton of normal RBCs, the average
end-to-end distance of the spectrins, i.e., the average distance between
nodes, is about $50-60$ $nm$. Here we summarize our experimental measurements
using Atomic Force Microscopy (AFM) and micropipette aspiration techniques.
The measurements show correlation between the stiffening of the iRBC
after chloroquine treatment as well as elongation of spectrins which
motivate our theoretical model. 

\begin{figure}
\centerline{\includegraphics[scale=0.55]{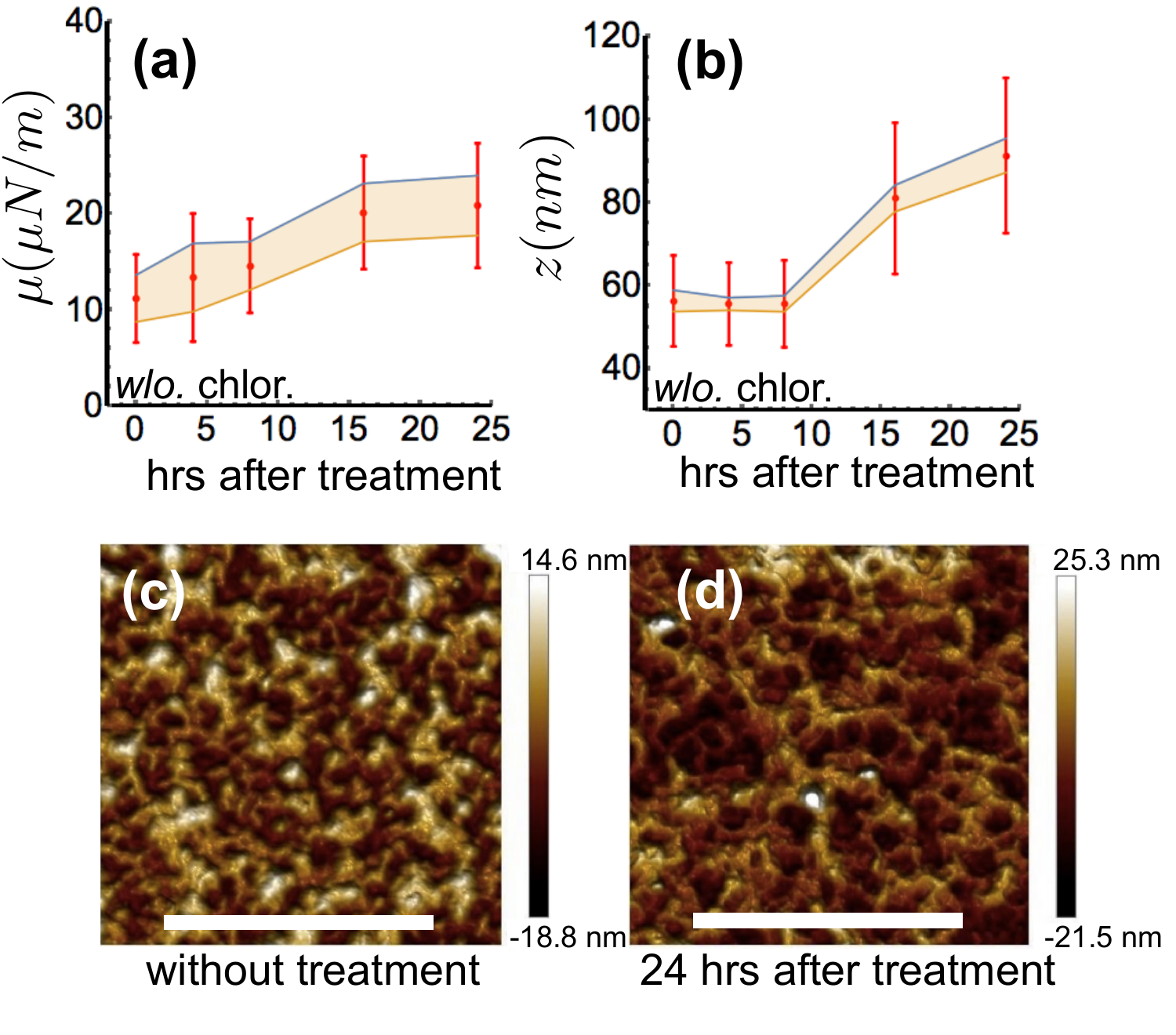}}

\protect\caption{\label{fig:Experimental-results}Experimental results. (a): Shear
modulus $\mu$ as a function of the time after chloroquine treatment.
(b): The end-to-end distance $z$ as a function of the time after
the treatment. In both (a) and (b), $0$ hour data corresponds to cells without chloroquine treatment. The shaded areas between the blue and orange curves in (a) and (b) indicate the $90\%$ point-wise confidence band for the mean values respectively. (c) and (d): Typical images from AFM at different stages
of the treatment (Without chloroquine treatment for (c) and 24 hrs after the treatment for (d)). The color indicates
the relative height of the corresponding point. The scale bar in (c)
and (d) is $1~\mu m$. }
\end{figure}

The shear modulus of the iRBC is measured by micropipette aspiration\cite{evans1973,hochmuth2000}.
Fig. \ref{fig:Experimental-results}(a) shows that the shear modulus
of iRBC increases by about a factor of $2$ after $24$ hours of chloroquine
treatment. The cytoskeleton structure is measured using AFM with a
similar method as that in \cite{shi13}. From the AFM images, we measured
the end-to-end distances of spectrins in the cytoskeleton network (Fig.
\ref{fig:Experimental-results} (b) and Fig. S3). The result shows an overall
increase of the end-to-end distances after chloroquine treatment.
The typical AFM images are shown in Fig. \ref{fig:Experimental-results}(c)
and (d). The larger dark spots in Fig. \ref{fig:Experimental-results}
(d) indicate larger holes in the cytoskeletal network, consistent
with our measurement of the end-to-end distances of spectrins.

\subsection{Numerical calculation\label{sub:Numerical-result}}

\begin{figure}
\centerline{\includegraphics[scale=0.3]{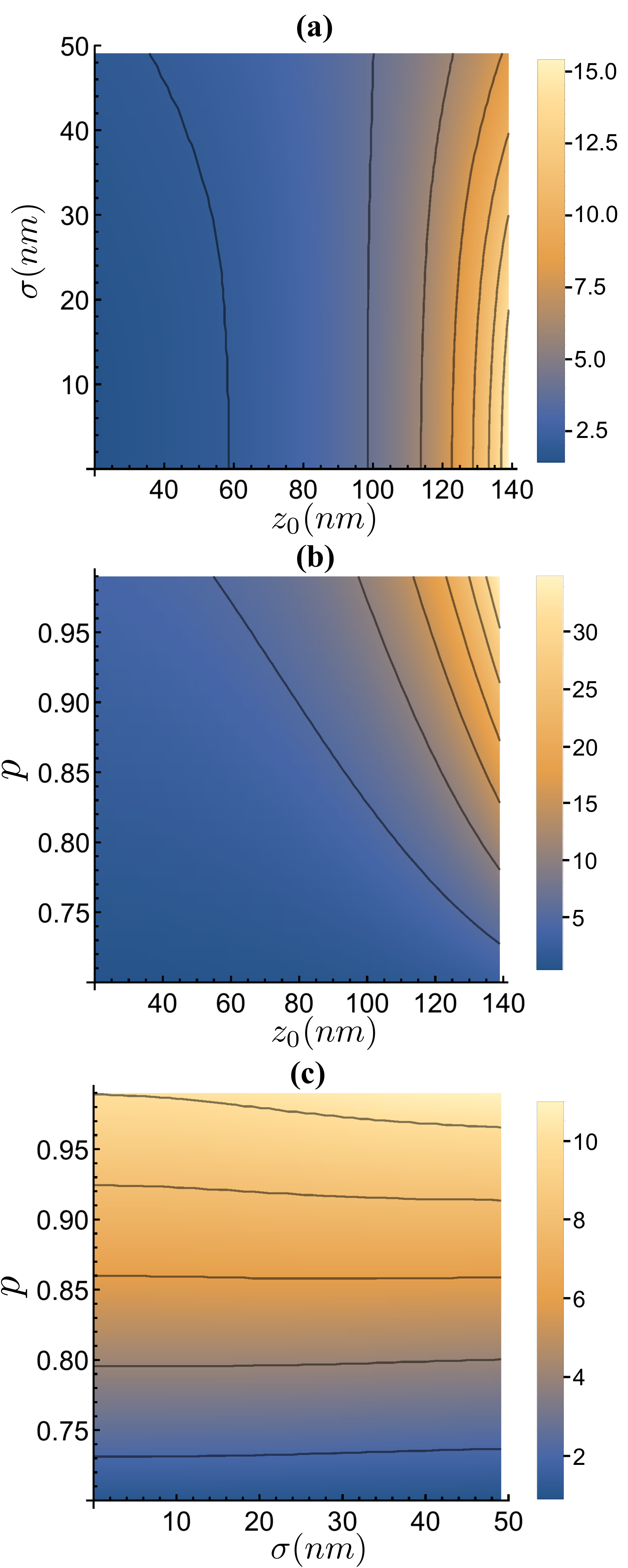}}

\protect\caption{\label{fig:Numerical-results}Numerical results for the effective
spring constant $k_{m}.$ (a) The dependence of $k_{m}$ on $z_{0}$
and $\sigma$ (Eq. \ref{eq:length-distribution}) with the
probability for one spectrin edge to present $p=0.8$ fixed. (b) The
dependence of $k_{m}$ on $z_{0}$ and $p$ with $\sigma=30~nm$ fixed.
(c) The dependence of $k_{m}$ on $\sigma$ and $p$ with $z_{0}=100~nm$
fixed. The color bars show the value of $k_{m}$ in the unit of $\mu N/m$.
The units match the units used in relevant experiments.}

\end{figure}

For a general probability density function ${\cal P}(k)$, Eq.
\ref{eq:EMT} can be solved numerically. Here we consider the variation
of the lengths of spectrins in the network, as well as the probability
of missing spectrins between nodes. Because the effective spring constant
of the spectrin depends on the end-to-end distance, the length distribution
introduced by Eq. \ref{eq:length-distribution} induces the
probability distribution of elastic constants for spectrins. Combined
with the probability of missing a spectrin connection ($1-p$), the probability density function of
the spring constants in the cytoskeletal network can be written as
\begin{equation}
{\cal P}(k)=p*\tilde{{\cal P}}(z)\frac{dz}{dk}+(1-p)\delta(k)\label{eq:distribution-general}
\end{equation}
where $\tilde{{\cal P}}(z)$ is the distribution of the end-to-end distance $z$ and normalized
for $0<z<L$, i.e., $\int_{0}^{L}\tilde{{\cal P}}(z)dz=1$, and $\delta(k)$ is the Dirac Delta function. In the
first term on the right-hand-side, $\tilde{{\cal P}}(z)\frac{dz}{dk}$
gives the distribution of the spring constant $k$. Using the probability
density function (eq. \ref{eq:distribution-general}) the EMT equation
(Eq. \ref{eq:EMT}) is re-written as
\begin{equation}
\int\frac{\tilde{{\cal P}}(z)}{1-\alpha(1-k(z)/k_{m})}dz=(1-\frac{1-p}{1-\alpha})/p\label{eq:EMT-different-length}
\end{equation}
As discussed earlier, the value of $\alpha$ depends on the geometry
of and the pre-stresses in the network. Here we take the value $\alpha=2/3$
for stress-free triangular network \cite{feng85} and solve $k_{m}$ numerically.
Tang $et.al.$ \cite{tang88} shows the relation between $\alpha$
and pre-strain in the network. We found that changing the value of
$\alpha$ (from $1/3$ to $2/3$) changes the value of $k_m$ but does not change the qualitative dependence of $k_m$ on $p$, $z_0$, and $\sigma$ as shown in Fig. \ref{fig:Numerical-results}. Once we obtain
the effective spring constant $k_{m}$ for the equivalent regular
network, the in-plane shear modulus $\mu$, to the lowest order, is
linearly proportional to $k_{m}$. Hence, to study how $\mu$ depends on the
disorders, it is equivalent to find out the dependence of $k_{m}$ on the connection
formation probability $p$, the average end-to-end distance $z_{0}$,
and the variance $\sigma^{2}$ of $z$. 

Our numerical results are summarized in Fig. \ref{fig:Numerical-results}.
With fixed value of $p=0.8$ (Fig. \ref{fig:Numerical-results}(a)),
the effective spring constant $k_{m}$ increases with $\sigma$ for
relatively small values of $z_{0}$, but decreases with $\sigma$
for relatively large values of $z_{0}$ ($z_{0}\apprge100~nm$). On
the other hand, $k_{m}$ always increases with increasing $z_{0}$
given the value of $\sigma$. This supports the prediction that the
experimentally observed cell stiffening is related to the observed
lengthening of spectrins in the cytoskeleton network. With fixed value
of $\sigma=30~nm$ (Fig. \ref{fig:Numerical-results}(b)), as one
may expect, $k_{m}$ increases with increasing $z_{0}$ given the
value of $p$, and also increases with increasing $p$ given the 
value of $z_{0}$. With fixed value of $z_{0}=100~nm$ (Fig. \ref{fig:Numerical-results}(c)),
$k_{m}$ increases with increasing $p$ given the value of $\sigma$.
But similar as shown in Fig. \ref{fig:Numerical-results}(a), $k_{m}$
increases with increasing $\sigma$ for relatively high values of $p$
($p\apprge0.86$) while it decreases with increasing $\sigma$ for
relatively low values of $p$. Therefore, our numerical results suggest
that, given the value of $p$ (e.g., Fig. \ref{fig:Numerical-results}(a)),
for relatively ``strong'' network (large values of $k_{m}$), increasing
the variance $\sigma^{2}$ of $z$ tends
to weaken the network, while for relatively ``weak'' network (small
value of $k_{m}$), increasing the variance $\sigma^{2}$ tends to
stiffen the network. However, given a relatively medium value of $z_{0}$
(e.g., $z_{0}=100~nm$ in Fig. \ref{fig:Numerical-results}(c)),
increasing $\sigma$ stiffens relatively ``strong'' network, but
weakens relatively ``weak'' network. This nontrivial dependence
of $k_{m}$ on $\sigma$ may be utilized in cellular functions or
material designs.

\subsection{Analytical solutions}

Here we adopt analytical solutions from previous studies (e.g., \cite{feng85,mao13})
but replace the harmonic springs with worm-like chains. To make the
model traceable, we only consider the deviation from the regular triangular
network caused by missing spectrin connections and neglect the length
heterogeneity. Aiming at the dependence of shear modulus on the
average spectrin length, here we will establish a correlation between
the experimentally observed spectrin lengthening and cell stiffening
phenomena. In this case, the distribution function takes the following
form \cite{feng85,sheinman12} 
\[
{\cal P}(k'\text{)=}p\delta(k'-k(z))+(1-p)\delta(k')
\]
Different from purely linear springs, stretching spectrins not only
introduces internal pre-stresses to the network, but also increases
the effective spring constants $k(z)$ of individual spectrins. If
we use the average extension $<z>$ to replace the inhomogeneous end-to-end
distance $z$, Eq. \ref{eq:EMT} can be solved exactly \cite{feng85}.
Again, the length-dependent spring constant only sets up the initial
value of $k(z)$. After that, we treat all the spectrins as Hookean
springs and $k$ as independent of $z$, and only consider
the linear response of the network for small perturbations. With this
consideration, Eq. \ref{eq:EMT} is solved to give us:

\begin{equation}
k_{m}=k(z)\frac{p-p_{c}}{1-p_{c}}=\frac{1}{\beta l_{p}L}(1+\frac{1}{2}(1-\frac{z}{L})^{-3})\frac{p-p_{c}}{1-p_{c}}\label{eq:effective-spring-constant}
\end{equation}
Here we consider $p$ as a constant independent of initial stresses.
The case of a stress-dependent $p$ will be discussed in the Discussion
section. Thus, we map the disordered
network to a regular triangular network consisting of springs with a 
single spring constant $k_{m}$. It is clear that, in our model, cell
stiffening comes directly from the nonlinearity of the WLC model,
as shown by the $z$-dependent term on the right hand side of Eq. \ref{eq:effective-spring-constant}. The macroscopic in-plane shear
modulus $\mu$ of the 2D network is linearly proportional to the effective
spring constant, i.e., $\mu\propto k_{m}$.

On the other hand, it is known that pre-stresses can stiffen the network
(e.g., \cite{tang88,boal98,sheinman12}), a similar effect as discussed
in subsection Numerical calculation. In a cellular environment,
the pre-stresses can be caused by various processes, e.g., the expression
and export of transmembrane proteins onto the iRBC membrane that bind
to spectrins, the formation of ``knobs'' at dispersed sites
on the RBC membrane that can link to near-by spectrins, the expansion
of the cell's volume due to the invaded and multiplying malarial parasites
and so on. In our model, without knowing more details about these
intracellular processes, we use a single stress-dependent pre-factor
$a(P)$ to account for the influence of the pre-stresses on the elastic
properties of the cytoskeletal network. With that in mind, the shear
modulus of the 2D network is given by 
\begin{equation}
\mu=a(P)k_{m}=\frac{a(P)}{\beta l_{p}L}(1+\frac{1}{2}(1-\frac{z}{L})^{-3})\frac{p-p_{c}}{1-p_{c}}\label{eq:analytical-shear-modulus}
\end{equation}
where $P$ is the pre-stress in the network, defined to be negative
for networks under tension. 

\begin{figure}
\centerline{\includegraphics[scale=0.25]{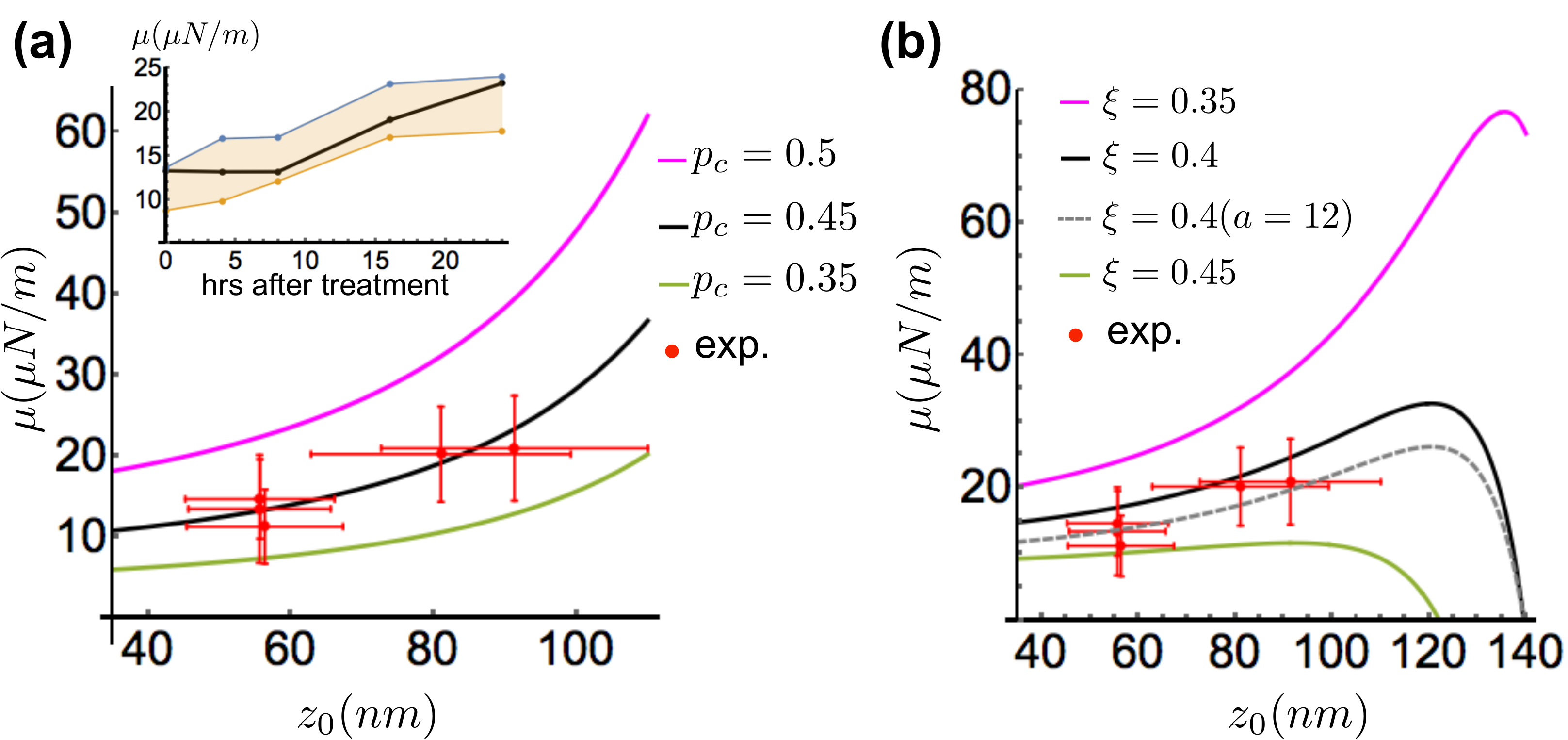}}

\protect\caption{\label{fig:Analytical-results}Analytical results and comparison with
experiments. (a) Dependence of the shear modulus $\mu$ on the average
end-to-end distance $z_{0}$. Red dots with error bars are experimental
data. Different curves correspond to different values of $p_{c}$. Inset: comparison between the analytical result and the experiments. The black curve is the analytical result, taking the same parameters as the black curve in (a) ($p_c=0.45$). The shaded area indicates the $90\%$ point-wise confidence band for the experimentally obtained mean values of the shear modulus. (b) Non-monotonic dependence of $\mu$ on $z_{0}$ for different values of $\xi$ (Eq. \ref{eq:kinetic-binding}),
when the unbinding kinetics is taken into account.}

\end{figure}

Fig. \ref{fig:Analytical-results}(a) shows our analytical results
for different values of $p_{c}$. We focus on the dependence of the
shear modulus $\mu$ on the average end-to-end distance $z_{0}$ and
we use $p=0.55$ that is approximated from our AFM measurements.
Qualitatively, it is clear that the shear modulus of the network increases
with increasing average end-to-end length $z_{0}$ of the spectrins in the network.
By adjusting the value of $p_{c}$ and $a(P)$, our results agree
with experiments (red dots with error bars in Fig. \ref{fig:Analytical-results}(a))
well. The comparison with experiments is elaborated below.

\subsection{Comparison with experiments}

Eq. \ref{eq:analytical-shear-modulus} is used to compare with
experiments. By using micropipette aspiration techniques, we found
that the malaria iRBCs become stiffer after chloroquine treatment
(Fig. \ref{fig:Experimental-results} (a)). In the meantime, through
AFM measurements, we found that the average end-to-end distance of
spectrins in the cytoskeleton increases after the drug treatment (Fig.
\ref{fig:Experimental-results} (b)). This coincidence is also noticed
in iRBCs at different stages of malaria infection \cite{chien1978,nash89,paulitschke1993,glenister2002,shi13}.
In our experiments, both the shear modulus $\mu$ and the spectrin
lengths were measured (Fig. \ref{fig:Analytical-results}(a)), and
the data showed a positive correlation between these two quantities.
In the meantime, the average connectivity is also obtained,
which is around $3.3$ across several samples. So we use this number
to determine the value of $p\approx3.3/6=0.55$ in our model. For
other parameters, we use $L=200~nm$, $T=300~K$ and $l_{p}=10~nm$,
which agree with the values reported in the experiments or literatures
(\cite{stokke85,svoboda92,li05}). $a$ and $p_{c}$ are used as fitting
parameters. Using the values $a=15$ and $p_{c}=0.45$, our results
(Eq. \ref{eq:analytical-shear-modulus}) agree well with the
experimental observations (Fig. \ref{fig:Analytical-results}(a) and the inset). For the black curve in the inset, the mean values of the lengths of spectrins at different times are used in Eq. \ref{eq:analytical-shear-modulus} to obtain the values of $\mu$ respectively. 
Due to the relatively large error bars in the experiments, the values of $a$ and $p_c$ can vary in a certain range. Better estimations of the parameter values are worthy of further investigations and more precise experimental measurements in our future work. However, the agreement supports the prediction that the shear modulus of the network is largely affected by the average length of edges (e.g., spectrins) in the cytoskeletal network.

\section{DISCUSSION \label{sec:Discussion}}

Several questions deserve further discussions here. Firstly, the binding/unbinding
kinetics between spectrins and the nodes (protein complexes) can be
force-dependent. Here we simply assume an exponential dependence of
the edge missing probability on the force, i.e., $(1-p)\sim\exp(\beta x^{\#}f)$,
where $f=f(z)$ depends on the end-to-end distances of the spectrins
as in Eq. \ref{eq:wlc-force-extension}. By plugging in $p=1-\xi\exp(\beta x^{\#}f(z))$
in Eq. \ref{eq:analytical-shear-modulus}, we obtain:
\begin{equation}
\mu=\frac{a(P)}{\beta l_{p}L}(1+\frac{1}{2}(1-\frac{z}{L})^{-3})\frac{(1-\xi\exp(\beta x^{\#}f(z)))-p_{c}}{1-p_{c}}\label{eq:kinetic-binding}
\end{equation}
where $x^{\#}$ is a parameter describing the length scale between
the binding potential minimum and maximum. In Eq. \ref{eq:kinetic-binding}, $z$ is approximated as the average end-to-end distance $<z>$. Two competing effects are
immediately noticed from Eq. \ref{eq:kinetic-binding}. The
$(1+\frac{1}{2}(1-\frac{z}{L})^{-3})$ term coming from the WLC model
predicts cell stiffening due to lengthening of spectrins, while the
other term $1-\xi\exp(\beta x^{\#}f(z))$ weakens the network. For
relatively short spectrin lengths, the cell stiffens due to the stiffenings
of spectrins as their lengths are increased. However, at large
spectrin lengths that may be beyond the measurements of current experiments,
the weakening term dictates the elastic properties of the network
and the bond forming probability $p$ may even drop below the rigidity
percolation threshold $p_{c}$. Thus a turning point should be observed
if the length of the spectrins keep increasing, dividing the two different
dependences of shear modulus on $z$. Fig. \ref{fig:Analytical-results}(b)
confirms this prediction, and shows the dependence of the shear modulus
$\mu$ on the average end-to-end distance $z_{0}$ of the spectrins
for different values of $\xi$. In Fig. \ref{fig:Analytical-results}(b),
we use $x^{\#}=1~nm$ that is in the range of values for similar problems\cite{efremov11}
and $p_{c}=0.45$. Other parameters take the same values as those
for Fig. \ref{fig:Analytical-results}(a). The curves in Fig. \ref{fig:Analytical-results}(b) are slightly shifted from those in Fig. \ref{fig:Analytical-results}(a) due to the differences between Eq. \ref{eq:analytical-shear-modulus} and Eq. \ref{eq:kinetic-binding}. The agreement with experiments (red dots with error bars) can be retained by only slightly changing the value of $a$ ($a=12$ for the gray dashed curve in Fig. \ref{fig:Analytical-results}(b)) with other parameters stay the same. Having that said, here we are more interested in the qualitative non-monotonic dependence of $\mu$
on $z_{0}$, while the exact shape of the curve needs to be further determined
by parameters from future experimental investigations. 

Secondly, it is noticed that pre-stress can significantly alter the
geometry or topology of the network through stretching of spectrins. 
The effects of pre-stresses in cells and cytoskeletons have been studied extensively previously regarding the stability of the triangular cytoskeletal network and the origin of the stresses are worthy of further studies. One scenario is the stress generated in the cytoskeleton of malaria infected RBCs. It is known that after the malaria infection,
``knobs'' are expressed across the membrane, which are complexes
of malarial parasite exported proteins. These knobs are found capable
of binding to spectrins and can possibly introduce stresses in the
cytoskeletal network\cite{shi13}. A different scenario is the iRBCs
after chloroquine treatment. In this case, the oxidative stress resulted
from the drug treatment may be responsible for the change of mechanical
properties of the cytoskeleton \cite{hale11}. We suspect that this
oxidative stress may be related to the formation of new protein complexes
that binds to spectrins, and subsequently generate stresses in the
network. In this case, the effect of chloroquine treatment is partially
impaired by the continuous stiffening of the cells, because stiffer
cells are sometimes easier to block the blood circulation. Further
experimental investigations of the origin of the stresses are our main focus in the future and  will improve our understanding of the mechanism of fatal diseases and drug treatments. 

From a broader perspective, although here we looked at the 2-dimensional
network, especially the RBC cytoskeleton as an example, our model
in general bridges the properties of biopolymers at the molecular
level and the elastic properties at the cellular level. Similar models
may be extended to other dimensions and shapes. Our results also show
a possible mechanism for cells to adjust their stiffness from the
molecular level and may stimulate new ideas in material designs.

\section{CONCLUSION}

To conclude, we investigated the relationship between the network
structure and its stiffness. We focused on how the disorders introduced
to a regular 2D triangular network affect its macroscopic in-plane
shear modulus. By applying the worm-like chain model to single spectrins
that form the edges of the network, we found that the shear modulus
of the network increases with the average length of the spectrins.
Our result agrees with previous studies in that removing some edges
with probability $1-p$ weakens the network, which shows a competing
effect with the stiffening caused by lengthening of spectrins (worm-like chains). On the other hand, as discussed in the results section, depending on the values of the average length $z_{0}$ or $p$, the shear modulus may increase or decrease
with increasing variance ($\sigma^{2}$) of the length distribution (Fig. \ref{fig:Numerical-results}(a)(c)). Furthermore, if the stress-dependent binding/unbinding kinetics is taken into account, the shear modulus has a first-increase-then-decrease behavior as the
average length of the spectrins increases. More importantly, our results
agree well with recent experimental observations for malaria infected
red blood cells (iRBCs), which shows a correlation between the lengthening
of spectrins (edges of the cytoskeleton network) and stiffening of
the cells (increase of shear modulus). Further investigations of what
causes the lengthening of spectrins or the internal stresses in the
cytoskeleton will be valuable in identifying new treatment methods
or targeted therapy for RBC-related diseases, such as malaria infection. 

\section*{AUTHOR CONTRIBUTIONS}
LL, XX, CTL, and JC designed the research. LL and XX performed the research. LL, CTL, and JC wrote the paper.

\section*{ACKNOWLEDGMENTS}
We acknowledge the financial assistance of Singapore-MIT Alliance
for Research and Technology (SMART) and National Science Foundation
(NSF) (CHE\textendash 1112825).

%\bibliographystyle{apsrev4-1}
%\bibliography{rbc}

%%%%%%%%%% Merge with supplemental materials %%%%%%%%%%
\pagebreak
\widetext
%%%%%%%%%% Merge with supplemental materials %%%%%%%%%%
%%%%%%%%%% Prefix a "S" to all equations, figures, tables and reset the counter %%%%%%%%%%
\setcounter{equation}{0}
\setcounter{figure}{0}
\setcounter{table}{0}
\setcounter{page}{1}
\makeatletter
\renewcommand{\theequation}{S\arabic{equation}}
\renewcommand{\thefigure}{S\arabic{figure}}
\renewcommand{\bibnumfmt}[1]{[S#1]}
\renewcommand{\citenumfont}[1]{S#1}
%%%%%%%%%% Prefix a "S" to all equations, figures, tables and reset the counter %%%%%%%%%%
\section*{SUPPORTING FIGURES}
\begin{figure}[h]
\centerline{\includegraphics[scale=0.8]{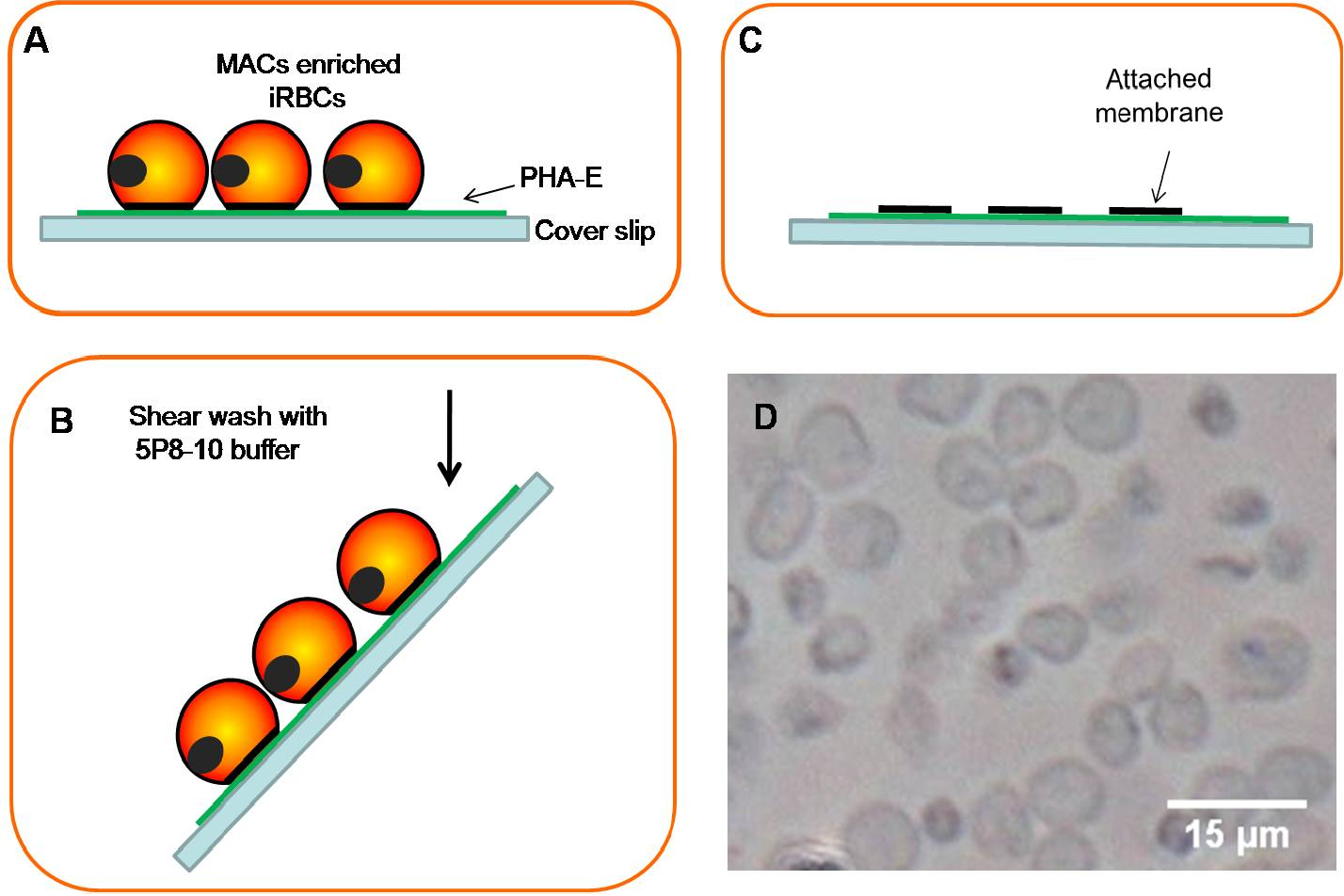}}

\protect\caption{\label{fig:s1} Cytoplasmic-surface-exposed membrane preparation process. A. MACs enriched iRBCs was first adhered to a PHA-E coated cover slip. B. Shear washing of adhered iRBCs using the 5P8-10 buffer at an angle of around 20$^\circ$ by syringe to peel off the top iRBC membrane. C. Cytoplasmic-surface-exposed membrane remained on the cover slip after shear washing. D. Phase contrast imaging of iRBC cytoplasmic-surface-exposed-samples.}
\end{figure}

\begin{figure}[h]
\centerline{\includegraphics[scale=0.8]{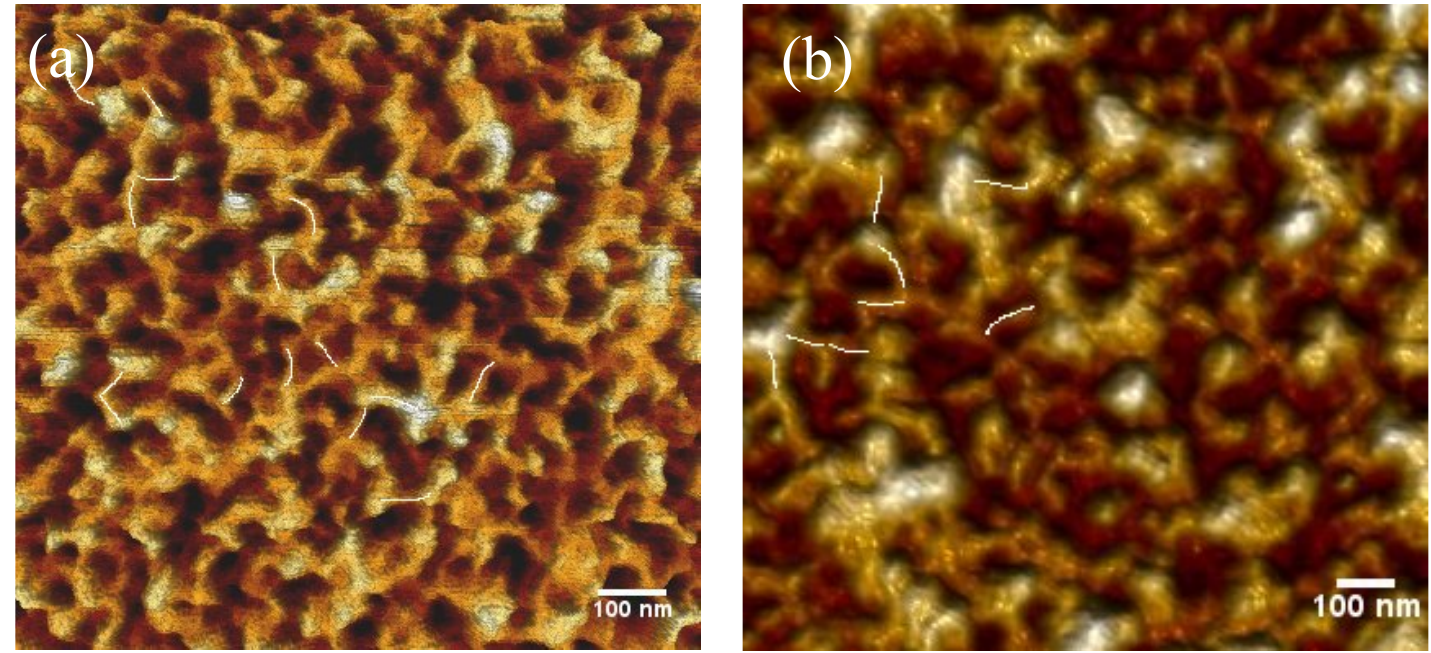}}

\protect\caption{\label{fig:s2} (a): 3D structure spectrin network constructed by JPKPM Data processing software for AFM image obtained using JPK Nanowizard\textregistered AFM. (b): 3D structure spectrin network constructed by using NanoScope Analysis 1.5 software for AFM images obtained using Dimension FastScanTM Atomic Force Microscope.}
\end{figure}

\begin{figure}[h]
\centerline{\includegraphics[scale=1.2]{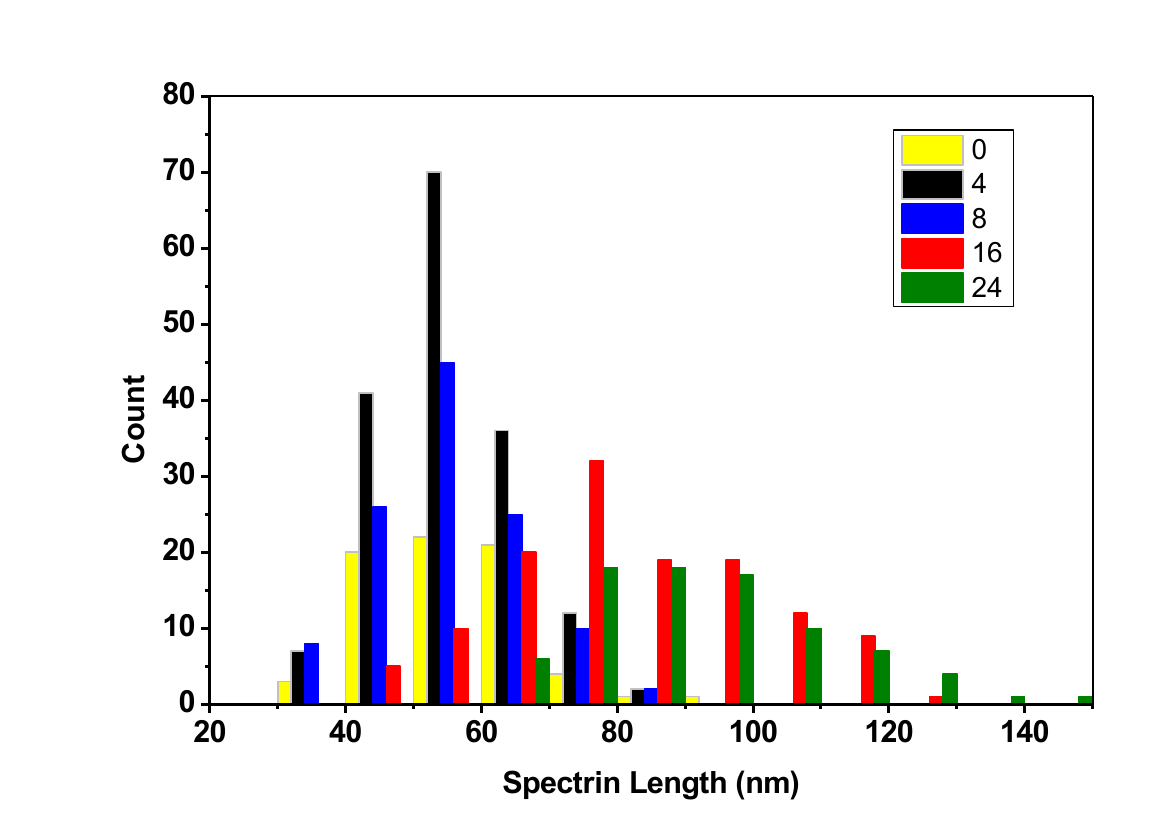}}

\protect\caption{\label{fig:s3} Distribution of spectrin lengths measured after different durations of chloroquine treatment. The time durations of treatment are shown in the legend.}
\end{figure}

\end{document}